
%
%

\magnification=\magstep2

\hsize=6.45truein
\vsize=8.89truein

\pageno 1

\baselineskip=16.7pt

\def\vecS{{\vec S}}

\noindent
\centerline{\bf Impurity-Bond Effect on the Spin-1 Antiferromagnetic}

\centerline{\bf Heisenberg Chain}

\vskip 18pt

\centerline{Makoto K{\sevenrm ABURAGI}, Isao H{\sevenrm ARADA}$^{\dagger}$
and Takashi T{\sevenrm ONEGAWA}$^{\dagger\dagger}$}

\vskip 18pt

\centerline{\it Department of Informatics, Faculty of Cross-Cultural Studies,}

\centerline{\it Kobe University,}

\centerline{\it Tsurukabuto, Nada, Kobe 657}

\centerline{$^{\dagger}${\it Department of Physics, Faculty of Science,
Okayama University,}}

\centerline{\it Tsushima, Okayama 700}

\centerline{$^{\dagger\dagger}$\it Department of Physics, Faculty of Science,
Kobe University,}

\centerline{\it Rokkodai, Kobe 657}

\vskip 18pt

\centerline{(Received December 24, 1992)}

\vskip 18pt

\parindent=1.5pc
Low-lying excited states of the spin-1 antiferromagnetic Heisenberg
chain with an impurity bond are investigated in terms of domain-wall
excitations by means of analytical methods as well as a method of
numerical diagonalization.  It is shown that 1) the impurity bond
brings about a massive triplet mode in the Haldane gap, 2) the triplet
state comprises three of the four ground states of an open chain, 3) the
excited states with the domain wall trapped at the impurity bond, including
the massive triplet mode, exhibit a N\'eel-type configuration of the magnetic
moment around the impurity bond. On the basis of the present results, we
propose a new energy versus magnetic-field diagram.

\vfill
\eject
%
%




\baselineskip=16.7pt

\def\vecS{{\vec S}}
\def\lsim{\,$\raise0.3ex\hbox{$<$}\llap{\lower0.8ex\hbox{$\sim$}}$\,}

\parindent=1.5pc
The spin-1 antiferromagnetic Heisenberg chain has been the subject of a large
number of theoretical and experimental studies since Haldane's
prediction$^{1)}$ of the difference between integer-spin and
half-integer-spin antiferromagnetic Heisenberg chains.  Most of the theoretical
studies done so far have been concerned with the properties of
the periodic chain.
Recently Kennedy$^{2)}$ found that the open chain has a fourfold degenerate
ground state composed of a singlet and a triplet which we call the Kennedy
triplet, in contrast to a unique singlet ground state of the periodic
chain.$^{3)}$  The fourfold degeneracy of the ground state, which was
originally found in the so-called AKLT model$^{3)}$ with open boundary
conditions, is considered to reflect the hidden $Z_2\!\times\!Z_2$ symmetry in
the open chain.$^{4)}$

\parindent=1.5pc
In this letter we investigate low-lying excited states which include the
Kennedy triplet of the spin-1 antiferromagnetic Heisenberg chain with an
impurity bond.  The impurity bond plays a mediating role between the open
chain and the periodic chain.  Using analytical methods as well as a method of
numerical diagonalization, we calculate the energies of the low-lying excited
states and also the magnetic moment at each site of the Kennedy triplet.  In
particular, exploring the dependence on the impurity-bond strength of the
energy difference between the Kennedy triplet and the singlet is interesting
and important, since it yields information on the transition of the ground
state from the fourfold to the unique one.

\parindent=1.5pc
Before going into the details of the calculation, we briefly discuss our
physical picture of the impurity-bond effect.  It is convenient in many cases
to express elementary excitations in the antiferromagnetic Heisenberg chain in
terms of domain-wall excitations.
If the excited state of the spin-1 periodic chain is the sliding mode of
domain-wall excitations, it is expected,
analogously to the spin-1/2 Ising-like periodic chain,$^{5)}$ that the
impurity bond traps the domain wall and brings about the massive triplet mode
below the bottom of the energy continuum.
For the open chain, this massive mode
should comprise three of the four ground states.  It is also expected that the
excited states with the trapped domain wall, including the massive triplet
mode, exhibit a N\'eel-type configuration of the magnetic moment around the
impurity bond.

\parindent=1.5pc
We express the Hamiltonian which describes the spin-1 antiferromagnetic
Heisenberg chain with an impurity bond as
$$ \eqalignno{
   {\cal H}& = \sum_{\ell=1}^{N-1} h_{\ell,\ell+1} + \kappa\, h_{N,1}
            + D \sum_{\ell=1}^{N}\bigl(S_\ell^z\bigr)^2 ~, & (1{\rm a})  \cr
 & h_{\ell,\ell'}
     = 2\bigl(S_\ell^x S_{\ell'}^x + S_\ell^y S_{\ell'}^y
     + \lambda S_\ell^z S_{\ell'}^z\bigr)
     + 2K\bigl(\vecS_\ell\cdot\vecS_{\ell'}\bigr)^2  \qquad (\lambda>0)~,~~~~~
                                                             & (1{\rm b})  \cr}
$$
where $\kappa$ represents the strength of the impurity bond,
with $\kappa\!=\!0$ and
$\kappa\!=\!1$ corresponding, respectively, to the open chain and the periodic
chain.  For convenience, we denote as $E_{\rm s}$ the energy of the state
corresponding to the singlet ground state at $\lambda\!=\!\kappa\!=1$ and
$D\!=\!K\!=\!0$, and also as $E_{\rm t}^{(\tau)}$ the energies of the states
corresponding to the Kennedy triplet at $\lambda\!=\!1$ and
$\kappa\!=D\!=\!K\!=\!0$, where $\tau\!=\!0$ for the
$S_{\rm total}^z\!\equiv\!\sum_{\ell=1}^N\!S_\ell^z\!=\!0$ state and
$\tau\!=\!\pm$ for the $S_{\rm total}^z\!=\!\pm1$ states.  Furthermore, we set
$\Delta_0(\kappa)\!\equiv\!E_{\rm t}^{(0)}\!-\!E_{\rm s}$ and
$\Delta_1(\kappa)\!\equiv\!E_{\rm t}^{(\pm)}\!-\!E_{\rm s}$.

\parindent=1.5pc
We first calculate, by means of a variational method,
$E_{\rm s}$ and $\Delta_i(\kappa)$
($i\!=\!0$, $1$) for the case of $|\kappa|\!\!\ll\!\!1$.  We introduce the
following matrix-product-form wave functions $\Phi$, $\Phi_n^{(+)}$,
$\Phi_n^{(0)}$, and $\Phi_n^{(-)}$, generalizing
that of Kl\"umper {\it et al.}$^{6)}$ which was originally proposed by
Fannes {\it et al.}$^{7)\;}$  The function $\Phi$ describes the state with no
domain wall and is defined as
$$ \eqalignno{
  \Phi =&\, {\rm Trace}\bigl[\phi_1 \phi_2 \cdots \phi_{N-1} \phi_N \bigr]~,
                                                             & (2)  \cr
  & \phi_{\ell}
     = \cos(\theta_\ell)\,\zeta_\ell\,\sigma_z
         + {\sin(\theta_\ell)\over\sqrt{2}}
            \bigr(\alpha_\ell\,\sigma_+ + \beta_\ell\,\sigma_-\bigl)~,
                                                             & (3)  \cr}
$$
where $\alpha_\ell$, $\zeta_\ell$, and $\beta_\ell$ are the spin states
at the $\ell$-th site, which
correspond, respectively, to $S_\ell^z\!=1$, $0$, and $-1$, and
$\sigma_{\pm}\bigl[=\!(\sigma_x\pm i\sigma_y)/\sqrt{2}\,\bigr]$,
$\sigma_x$, $\sigma_y$, and $\sigma_z$ are the Pauli matrices.  On the other
hand, the function $\Phi_n^{(\tau)}$ describes the states
with a domain wall at the $n$-th bond between the $n$th and
$(n\!+\!1)$th sites,
and is defined in terms of $\phi_\ell$ and an operator $w$, which may be
called the wall operator, as
$$ \Phi_n^{(\tau)}
     = {\rm Trace}\bigl[\phi_1 \phi_2 \cdots \phi_n\,w\,\phi_{n+1}
                                    \cdots \phi_{N-1} \phi_N \bigr]~,
                                                               \eqno  (4)  $$
where $w\!=\!-\sigma_-$ for $\tau\!=\!+$, $w\!=\!\sigma_z$ for $\tau\!=\!0$,
and $w\!=\!\sigma_+$ for $\tau\!=\!-$.  It is noted that when $\lambda\!=\!1$,
$\Phi$ represents the singlet state, and $\Phi_n^{(+)}$, $\Phi_n^{(0)}$, and
$\Phi_n^{(-)}$ represent, respectively, the triplet states
with $S_{\rm total}^z\!=\!1$, $0$, and $-1$.  In the case of
$|\kappa|\!\ll\!1$ the domain wall is expected to be trapped at the impurity
bond, i.e., the $N$th bond.  We therefore use $\Phi$, $\Phi_N^{(+)}$,
$\Phi_N^{(0)}$, and $\Phi_N^{(-)}$ as trial functions.  For simplicity, we
assume that $\theta_\ell\!=\!\theta$ for all $\ell$'s.  Then, these trial
functions become equivalent to that used by Kennedy and Tasaki.$^{4)}$  We
have carried out a variational calculation with respect to $\theta$ for the
energy expectation values.  Up to the order of $N$, the four trial
functions lead
to the same solution $\theta\!=\!\tilde\theta$, where $\tilde\theta$ is given
by
$$  \tilde\theta
       = \cases{0 &
                        ~~~for~~~$D\ge{\rm Max}(4,\lambda)$~,             \cr
                {1\over 2}\arccos\Bigl({D-\lambda\over 4-\lambda}\Bigr) &
                        ~~~for~~~$4>D>2\lambda-4$~,                       \cr
                {\pi\over2} &
                        ~~~for~~~${\rm Min}(\lambda,2\lambda-4)\geq D$~.  \cr}
                                                         \eqno (5)
$$
This solution in the region $D\!\ge\!{\rm Max}(4,\lambda)$, that in the region
$4\!>\!D\!>\!2\lambda\!-\!4$, and that in the
region ${\rm Min}(\lambda,2\lambda-4)\!\geq\!D$ correspond, respectively, to
the\break large-$D$, Haldane, and N\'eel phases.  (In the calculation for the
Haldane phase we have neglected the factors $\cos^{N}\!\!\theta$ and
$\cos^{N}\!\!2\theta$.)  We denote $\Phi$ and $\Phi_N^{(\tau)}$ for
$\theta\!=\!\tilde\theta$ by $\tilde\Phi$ and ${\tilde\Phi}_N^{(\tau)}$,
respectively.  It is noted that $\tilde\Phi$ and ${\tilde\Phi}_N^{(\tau)}$
for $D\!=\!0$ and $\lambda\!=\!1$ are equivalent to the
valence-bond-solid$^{3)}$ wave functions expressed by the Schwinger
bosons.$^{8)}$

\parindent=1.5pc
Let us concentrate our attention on the Haldane phase.  For this phase, the
variational results for $E_{\rm s}$ and $\Delta_i(\kappa)$ are given by
$$ \eqalignno{
     E_{\rm s}
       &\simeq \langle\tilde\Phi|{\cal H}|\tilde\Phi\rangle        &     \cr
       &= - (N-1+\kappa)\Bigl\{\Bigl({4-D\over 4-\lambda}\Bigr)
                       \Bigl(1+{D\over 4}\Bigr) + 2K\Bigr\}
          + N {D(4-D)\over 2(4-\lambda)}~,~~~                    & (6)   \cr
     \Delta_0(\kappa) &\simeq \langle{\tilde\Phi}_N^{(0)}|{\cal H}
                                            |{\tilde\Phi}_N^{(0)}\rangle
                            - E_{\rm s}                                  \cr
       &= \kappa\,\Bigl\{{2(4-D)(4-2\lambda+D)\over(4-\lambda)^2}-K\Bigr\}
        \equiv \kappa\tilde\Delta_0~,                            & (7)   \cr
     \Delta_1(\kappa) &\simeq \langle{\tilde\Phi}_N^{(\pm)}|{\cal H}
                                            |{\tilde\Phi}_N^{(\pm)}\rangle
                            - E_{\rm s}                                  \cr
       &= \kappa\,{4-D\over 4(4-\lambda)^2}
              \bigl\{2(8+2D-\lambda D)-K(12-4\lambda+D)\bigr\}
        \equiv \kappa\tilde\Delta_1~.                            & (8)   \cr}
$$
When \hbox{$\kappa\!=\!0$}, as is expected, the four states $\tilde\Phi$ and
$\tilde\Phi_N^{(\tau)}$ are degenerate in energy, which reflects the hidden
\hbox{$Z_2\!\times\!Z_2$} symmetry in the open chain.  This degeneracy is
resolved linearly with respect to $\kappa$; in the case of
\hbox{$\kappa\!>\!0$}, either
$0\!<\!\Delta_1(\kappa)\!<\!\Delta_0(\kappa)$ or
$0\!<\!\Delta_0(\kappa)\!<\!\Delta_1(\kappa)$ depending on
whether $D(4\!+\!2\lambda\!-\!3K)$ is larger or smaller than
$4(4\!-\!K)(\lambda\!-\!1)$, and in the case of $\kappa\!<\!0$, either
$0\!>\!\Delta_1(\kappa)\!>\!\Delta_0(\kappa)$ or
$0\!>\!\Delta_0(\kappa)\!>\!\Delta_1(\kappa)$ depending on whether
$D(4\!+\!2\lambda\!-\!3K)$ is larger or smaller than
$4(4\!-\!K)(\lambda\!-\!1)$.  Thus, when $\lambda\!=\!1$ and $D\!=\!0$, a
transition of the ground state from a singlet to a triplet takes place at
$\kappa\!=\!0$ with increasing $\kappa$.  This finding indicates that the
impurity bond plays the role of the $Z_2\!\times\!Z_2$ symmetry breaking field
and that the $Z_2\!\times\!Z_2$ symmetry is not a characteristic property of
the bulk.

\parindent=1.5pc
The site dependence of the magnetic moments is calculated as
$$ \eqalignno{
     &\langle\tilde\Phi|S_\ell^z|\tilde\Phi\rangle
      = \langle{\tilde\Phi}_N^{(0)}|S_\ell^z|{\tilde\Phi}_N^{(0)}\rangle
      = 0 ~,                                               &  (9)  \cr
     &\langle{\tilde\Phi}_N^{(+)}|S_\ell^z|{\tilde\Phi}_N^{(+)}\rangle
      = -\langle{\tilde\Phi}_N^{(-)}|S_\ell^z|{\tilde\Phi}_N^{(-)}\rangle
      = \sin^2\!\tilde\theta\bigl(\cos^{\ell-1}\!2\tilde\theta
                               +\cos^{N-\ell}\!2\tilde\theta\bigr)~.~~~~~~~
                                                           & (10)  \cr}
$$
The result given by eq.$\,$(10), which is essentially the same as that
obtained by Kennedy and Tasaki,$^{4)}$ indicates that
$\langle{\tilde\Phi}_N^{(\pm)}|S_\ell^z|{\tilde\Phi}_N^{(\pm)}\rangle$ decays
exponentially with the decay constant $-(\ln|\cos2\tilde\theta\,|)^{-1}$ as a
function of the distance from the impurity bond.

\parindent=1.5pc
We briefly discuss the case in which the magnetic-field term
${\cal H}_{\rm mf}\!=\!-H_x S_{\rm total}^x\!-\!H_z S_{\rm total}^z$
$\bigl(S_{\rm total}^x\!\equiv\!\sum_{\ell=1}^N\!S_\ell^x\bigl)$ is added to
the Hamiltonian ${\cal H}$.  The energy difference $\Delta E$ (the energy
measured from $E_{\rm s}$) versus magnetic field $H_x$ or $H_z$ diagram can be
constructed by solving the equation,
$$  \bigl[\Delta_0(\kappa)-\Delta E\bigr]
    \bigl[\bigl\{\Delta_1(\kappa)-\Delta E\bigr\}^2-H_z^2\bigl]
      - G^2 H_x^2 \bigl\{\Delta_1(\kappa)-\Delta E\bigr\} = 0     \eqno (11)
$$
with
$$
  G = \langle{\tilde\Phi}_N^{(+)}|S_{\rm total}^x|{\tilde\Phi}_N^{(0)}\rangle
    = \langle{\tilde\Phi}_N^{(-)}|S_{\rm total}^x|{\tilde\Phi}_N^{(0)}\rangle
    = \sin 2\tilde\theta/(1 + \cos^2\!2\tilde\theta)~.          \eqno  (12)
$$
We note here that $\langle{\tilde\Phi}|S_{\rm
total}^x|{\tilde\Phi}_N^{(+)}\rangle\!=\!\langle{\tilde\Phi}|S_{\rm
total}^x|{\tilde\Phi}_N^{(0)}\rangle$ $=\!\langle{\tilde\Phi}|S_{\rm
total}^x|{\tilde\Phi}_N^{(-)}\rangle\!=\!\langle{\tilde\Phi}_N^{(+)}|S_{\rm
total}^x|{\tilde\Phi}_N^{(-)}\rangle\!=\!0$.  The explicit
results for the solution of eq.$\,$(11) will be discussed in a separate
paper.$^{9)}$

\parindent=1.5pc
Next we deal with the case of $\kappa\lsim 1$, again concentrating our
attention on the Haldane phase.  In this case it is important to take into
account the effect of the propagation of domain walls in the chain.  When
$\kappa\!\ne\!1$, the\break scattering of propagating domain walls by the
impurity bond produces a domain-wall bound state around it.  To discuss this
bound state, we employ Koster and Slater's method.$^{10)}$  (A more
sophisticated treatment will be discussed elsewhere.$^{11)}$)  This method
leads to the following result for the energy differences $\Delta_i(\kappa)$:
$$  \Delta_i(\kappa) \simeq \Delta_i(1) - a_i(1 - \kappa)^2~,    \eqno (13) $$
where $\Delta_i(1)$ and $a_i$ are constants, the latter being a quantity
related to the matrix elements of $h_{N,1}$, and are determined
phenomenologically below.  The present Koster-Slater analysis also leads to
the result that the energy difference between the bottom of the energy
continuum for $S_{\rm total}^z\!=\!0$ and the ground state and that
between the bottom of the energy continuum for
$S_{\rm total}^z\!=\!\pm1$ and the ground state are given, respectively, by
$\Delta_0(1)$ and $\Delta_1(1)$.  Accordingly, these energy differences are
independent of $\kappa$.

\parindent=1.5pc
We assume that eq.$\,$(13) also holds for $\kappa\!\sim\!0$ to determine
$\Delta_i(1)$ and $a_i$.  Then, $a_i$ should be equal to $\Delta_i(1)$, since
$\Delta_i(\kappa)$ should vanish at $\kappa\!=\!0$.  Thus, we have
$$  \Delta_i(\kappa) = \Delta_i(1)\Bigl\{1-(1-\kappa)^2\Big\}~.  \eqno (14) $$
Furthermore, if we use the variational results given by eqs.$\,$(7) and (8),
$\Delta_i(1)$ is determined to be $\Delta_i(1)\!=\!\tilde\Delta_i/2$.

\parindent=1.5pc
On the basis of the results of the above Koster-Slater analysis, we discuss how
the state at the bottom of the energy continuum changes with $\kappa$,
considering, for simplicity, the isotropic case of $\lambda\!=\!1$ and
$D\!=\!K\!=\!0$ in which
$\Delta_0(\kappa)\!=\!\Delta_1(\kappa)\!\equiv\!\Delta(\kappa)$.  At
$\kappa\!=\!1$ the energy difference between the bottom of the continuum for
the triplet state and that for the quintet state is
considered to be equal to $\Delta(1)$.  On the other hand, the binding energy
of one trapped domain wall at $\kappa\!=\!0$ is given by $\Delta(1)$ [see
eq.$\,$(14)], and the energy difference $\Delta(1)$ between the bottom of the
continuum for the triplet state and the ground state is independent of
$\kappa$.  Thus, when the value of $\kappa$ decreases, the state at the bottom
of the continuum is expected to change at $\kappa\!=\!0$ from the triplet to
the quintet, since at $\kappa\!\sim\!0$ the latter state is constructed by
adding one trapped domain wall to the former state.

\parindent=1.5pc
Combining these considerations with the variational result that
$\Delta(\kappa)$ is proportional to $\kappa$ at $\kappa\!\sim\!0$ [see
eqs.$\,$(7) and (8)] and also with the result given by eq.$\,$(14), we
conjecture that the energy (measured from $E_{\rm s}$) versus $\kappa$ diagram
is that which is schematically given in Fig.$\,$1.  Here we have assumed that
the energy difference between the bottom of the continuum and the ground state
remains unchanged also for \hbox{$\kappa\!<\!0$}.  According to this diagram
the Haldane gap is defined as the energy difference between the bottom of the
energy continuum and the ground state for any $\kappa$.

\parindent=1.5pc
In order to numerically examine the analytical results as well as the
conjecture discussed above, we have performed diagonalizations by
Lancz\"os method for
finite-$N$ chains in the isotropic case.  In Fig.$\,$2 we plot\break
$\Delta(\kappa)/\Delta(1)$ as a function of
$\kappa'\bigl[\equiv\!(\kappa\!-\!\kappa_{\rm c})/(1\!-\!\kappa_{\rm c})\bigr]$
for $N\!=\!12$ and $N\!=\!13$, where $\kappa_{\rm c}$ is the value of $\kappa$
at which $\Delta(\kappa)$ vanishes for each $N$.  The corresponding plot of
eq.$\,$(14), where $\kappa'\!=\!\kappa$, is also given in Fig.$\,$2.  From this
figure we see that the result for the $\kappa$ dependence of $\Delta(\kappa)$
obtained by the above single domain-wall analysis is in rather good agreement
with the numerical results.  In Fig.$\,$3 the energies (measured from
$E_{\rm s}$) of low-lying excited states for $N\!=\!13$ are plotted as a
function of $\kappa$; the dotted line is for the lowest triplet state [i.e.,
the dotted line shows $\Delta(\kappa)$], the dashed line is for the
second-lowest triplet state, and the dot-dashed line is for the lowest quintet
state.  This figure should be compared with Fig.$\,$1. We plot in Fig.$\,$4 the
$\ell$-dependence of the magnetic moments $\langle S_\ell^z\rangle$ of the
$S_{\rm total}^z\!=\!1$ state for the lowest triplet and of the
$S_{\rm total}^z\!=\!2$ state for the lowest quintet obtained for $N\!=\!13$
and $\kappa\!=\!0$ (the open chain), where $\langle\cdots\rangle$ stands for
the expectation value with respect to a given state.  This figure suggests a
N\'eel-type configuration of the magnetic moments around the impurity bond for
both states.  It should be noted that the magnetic moments shown in Fig.$\,$4
are quite different from $\langle S_\ell^z\rangle\!=\!S_{\rm total}^z/N$ for
$\kappa\!=\!1$ (the periodic chain) and that the values of the magnetic
moments at the edges of the open chain are about 60 percent of the value 1
of the full moment.  As may be seen from Fig.$\,$4, the decay constant of the
$S_{\rm total}^z\!=\!1$ state for the lowest triplet is much longer than the
variational result of $-(\ln|\cos2\tilde\theta\,|)^{-1}\!\sim\!0.9$.  In fact,
Miyashita and Yamamoto$^{12)}$ have very recently shown that the
$N\!\to\!\infty$ value of the decay constant of this state is given by about
6.  Details of the numerical calculation will be published in the near
future.$^{11)}$

\parindent=1.5pc
In conclusion, we emphasize that the Haldane gap should not be defined as the
singlet-triplet excitation energy, but as {\it the energy difference between
the bottom of the energy continuum and the ground state}.  We propose,
for the isotropic open chain, the energy (measured from $E_{\rm s}$) versus
magnetic-field $H_z$ diagram given by Fig.$\,$5 (see Fig.$\,$1 also).  It is
noted that the critical field $H^{\rm c}_z$, at which the finite magnetization
per site appears, is equal to $\Delta(1)$ as in the case of the isotropic
periodic chain.  Furthermore, we expect that for samples doped by nonmagnetic
impurities of a few or more percent, the magnetic response due to the
N\'eel-type configuration of magnetic moment near the edges can be observed
at a magnetic field even below $H^{\rm c}_z$.
This may be an important subject for
future experimental studies for examining our physical picture.

\parindent=1.5pc
The present work has been supported in part by a  Grant-in-Aid for Scientific
Research on Priority Areas, ^^ ^^ Computational Physics as a New Frontier in
Condensed Matter Research'', from the Ministry of Education, Science and
Culture.  One of the authors (M.~K.) gratefully acknowledges the support of
Fujitsu Limited.

\vfill\eject

\centerline{\bf References}

\item{1)} F.~D.~M.~Haldane:~Phys.~Lett.~{\bf 93A} (1983)
464;~Phys.~Rev.~Lett.~{\bf 50} (1983) 1153.

\item{2)} T.~Kennedy:~J.~Phys.~Condens.~Matter {\bf 2} (1990) 5737.

\item{3)} I.~Affleck, T.~Kennedy, E.~H.~Lieb and
H.~Tasaki:~Phys.~Rev.~Lett.~{\bf 59} (1987) 799;~Commun.~Math.~Phys.~{\bf 115}
(1988) 477.

\item{4)} T.~Kennedy and H.~Tasaki:~Commun.~Math.~Phys.~{\bf 147} (1992) 431.

\item{5)} T.~Moriwaki, M.~Kaburagi and T.~Tonegawa:~unpublished.

\item{6)} A.~Kl\"umper, A.~Schadschneider and J.~Zittartz:~J.~Phys.~A
{\bf 24} (1991) L955;~Z.~Phys.~B~{\bf 87} (1992) 281.

\item{7)} M.~Fannes, B.~Nachtergaele and R.~F.~Werner:~Europhys.~Lett.~{\bf 10}
(1989) 633; Commun.~Math.~Phys.~{\bf 144} (1992) 443.

\item{8)} D.~P.~Arovas and A.~Auerbach:~Phys.~Rev.~B {\bf 38} (1988) 316.

\item{9)} M.~Kaburagi and T.~Tonegawa:~in preparation.

\item{10)} G.~F.~Koster and J.~C.~Slater:~Phys.~Rev.~{\bf 96} (1954) 1208.

\item{11)} M.~Kaburagi:~in preparation.

\item{12)} S.~Miyashita and S.~Yamamoto:~preprint.

\vfill\eject

\centerline{\bf Figure Captions}

\vskip 9pt

{\leftskip=1.5pc
\parindent=-1.5pc
Fig.$\,$1.~~Schematic energy (measured from $E_{\rm s}$) versus $\kappa$
diagram for the isotropic case obtained by analytical methods.  The dotted
line is for the lowest triplet state and the dashed (dot-dashed) line is for
the bottom of the energy continuum consisting of triplet (quintet) states.
The gray region shows the energy continuum.\par}

\vskip 9pt

{\leftskip=1.5pc
\parindent=-1.5pc
Fig.$\,$2.~~Plots of $\Delta(\kappa)/\Delta(1)$ as a function of
$\kappa'\bigl[\equiv\!(\kappa\!-\!\kappa_{\rm c})/(1\!-\!\kappa_{\rm c})\bigr]$
for $N\!=\!12$ (open circles) and $N\!=\!13$ (open squares) in the
isotropic case, where $\kappa_{\rm c}$ is the value of $\kappa$ at which
$\Delta(\kappa)$ vanishes for each $N$.  The solid line is the corresponding
plot of eq.$\,$(14), where $\kappa'\!=\!\kappa$.\par}

\vskip 9pt

{\leftskip=1.5pc
\parindent=-1.5pc
Fig.$\,$3.~~Energy (measured from $E_{\rm s}$) versus $\kappa$ diagram for
$N\!=\!13$ in the isotropic case obtained by numerical diagonalization.  The
dotted line is for the lowest triplet state, the dashed line is for the
second-lowest triplet state, and the dot-dashed line is for the lowest quintet
state.\par}

\vskip 9pt

{\leftskip=1.5pc
\parindent=-1.5pc
Fig.$\,$4.~~Plots versus $\ell$ of the magnetic moments
$\langle S_\ell^z\rangle$  of the $S_{\rm total}^z\!=\!1$ state for the
lowest triplet (full circles) and of the $S_{\rm total}^z\!=\!2$ state for
the lowest quintet (open circles) obtained for $N\!=\!13$ and
$\kappa\!=\!0$.\par}

\vskip 9pt

{\leftskip=1.5pc
\parindent=-1.5pc
Fig.$\,$5.~~Energy (measured from $E_{\rm s}$) versus magnetic-field $H_z$
diagram proposed for the isotropic open chain.\par}

\bye

\bye